\documentstyle[11pt,epsfig,twoside]{article}
\newcommand{\mr}[1]{{\mathrm {#1}}}

\newcommand{\lm}{$\Lambda\;$}
\newcommand{\lme}{\Lambda}
\newcommand{\lmpdf}{$\Lambda^{\mr{pdf}}\;$}
\newcommand{\lmepdf}{\Lambda^{\mr{pdf}}\;}
\newcommand{\lmhsc}{$\Lambda^{\mr{hsc}}\;$}
\newcommand{\lmehsc}{\Lambda^{\mr{hsc}}\;}
\newcommand{\lms}{$\Lambda_{\overline{\mr{MS}}}\;$}
\newcommand{\lmse}{\Lambda_{\overline{\mr{MS}}}}

\newcommand{\bbare}[1]{\overline{\mr{#1}}}
\newcommand{\bbar}[1]{$\overline{\mr{#1}}$}
\newcommand{\reff}[1]{(\ref{#1})}
\newcommand{\be}{\begin{equation}}
\newcommand{\ee}{\end{equation}}
\newcommand{\bear}{\begin{eqnarray}}
\newcommand{\eear}{\end{eqnarray}}
\newcommand{\bc}{\begin{center}}
\newcommand{\ec}{\end{center}}
\newcommand{\bd}{\begin{description}}
\newcommand{\ed}{\end{description}}
\newcommand{\bit}{\begin{itemize}}
\newcommand{\eit}{\end{itemize}}
\newcommand{\ben}{\begin{enumerate}}
\newcommand{\een}{\end{enumerate}}

\begin{document}
\begin{flushright} PRA-HEP/97-7
\end{flushright}
\begin{center}
{\large \bf Jet rates in deep inelastic scattering: wherefrom comes the
sensitivity to $\alpha_s$?}

\vspace*{0.5cm}
{\em Ji\v{r}\'{\i} Ch\'{y}la and Ji\v{r}\'{\i} Rame\v{s}}

\vspace*{0.3cm}
Institute of Physics, Academy of Sciences of the Czech Republic,
Prague
\footnote{e--mail: chyla@fzu.cz, rames@fzu.cz}

\vspace*{0.4cm}
{\bf Abstract}
\end{center}

\noindent
For theoretically consistent
determination of $\alpha_s$ from jet rates in deep inelastic scattering
the dependence on $\alpha_s$ of parton distribution functions is in
principle as important as that of hard scattering cross--sections.
For the kinematical region accessible at HERA we investigate in detail
numerical importance of these two sources of the $\alpha_s$
dependence of jet rates.

\section{Introduction}
One of the problems of quantitative determination of the running of
$\alpha_s$ is related to the fact that it usually requires combining
results of different experiments in different kinematical regions.
Recently the
H1 \cite{H1} and ZEUS \cite{ZEUS} Collaborations have reported
evidence for the running of $\alpha_s(\mu)$ obtained from the
measurement of jet rates in
deep inelastic scattering (DIS) via the quantity
\be
R_{2+1}(Q^2)\equiv \frac{\sigma_{2+1}(Q^2)}
{\sigma_{1+1}(Q^2)+\sigma_{2+1}(Q^2)},
\label{R2+1}
\ee
where $\sigma_{k+1}$ denotes the cross--section for the production
of $k$ hard and one proton remnant jets. There are now several NLO
Monte--Carlo generators \cite{PROJET,MEPJET,DISJET},
suitable for analyses of
jet production in DIS. In this paper we shall concentrate on
the H1 analysis \cite{H1} which, using PROJET 4.1
\cite{PROJET} with the JADE
jet algorithm and
$y_c=0.02$, obtained the following result for
$\alpha_s(M_Z,\bbare{MS})$
\be
\alpha_s(M_Z,\bbare{MS})=
0.123\pm0.012(\mr{stat.})\pm0.008(\mr{syst.}).
\label{alphas}
\ee
In JADE cluster algorithm,
the jet resolution parameter $y_c$ is defined
as $y_c=(p_i +p_j)^2/W^2$, where $p_i,p_j$ are parton
momenta and $W^2=Q^2(1-x)/x$ stands for the square of the $\gamma$p CMS
energy. The result \reff{alphas} has been
extracted from the $Q^2$--dependence of $R_{2+1}(Q^2)$, displayed in
Fig. 2 of \cite{H1}, using only the two highest $Q^2$ data points.
Within the error bars the value \reff{alphas} is consistent with the
world average, but its accuracy is insufficient to draw
any firmer conclusions.
In the procedure adopted in \cite{H1} $\alpha_s(\mu,\bbare{MS})$
(or, equivalently, $\lmse$) was considered as a
free parameter in the hard scattering cross--sections, but not in the
parton distribution functions (PDF), for which the MRSH set was used.
However, as each set of PDF has a particular value of \lms built in,
one must check for the consistency between this input \lms and the
output one, obtained from comparison of \reff{R2+1} with data.  Clearly,
in a consistent determination of $\alpha_s$, \lms must be varied
simultaneously in PDF and parton level hard scattering cross--sections.
In practical applications to physical quantities
involving beside the hard scattering cross--sections also parton
distribution and/or fragmentation functions
it is often useful to know wherefrom comes most of the  sensitivity
to $\alpha_s$. The purpose of our paper is to investigate this question
in detail for the jet rates \reff{R2+1}.

Throughout this paper we stay in the conventional \bbar{MS}
renormalization scheme (RS) of the couplant $a\equiv\alpha_s(\mu)/\pi$
and omit therefore the specification ``\bbar{MS}'' in \lms as well as in
$\alpha_s(\mu,\bbare{MS})$.
The jet cross--sections $\sigma_{k+1}$ in \reff{R2+1} are given as
convolutions
\be
\sigma_{k+1}(Q^2,y_c,\lme)\equiv \sum_{i}\int^1_0 \mr{d}x
f_{i}(x,M,\lme)C_{k+1,i}(Q,M,x,y_c,\lme)
\label{sigmak+1}
\ee
of the parton level cross--sections $C_{k+1,i}$ and PDF
$f_i(x,M)$, evaluated
at the factorization scale $M$. The sum runs over
all parton species $i$. In perturbative QCD $C_{k+1,i}$
are given as expansions in the couplant
$a(\mu/\lme)$, taken at the hard scattering scattering scale
\footnote{Although in general $\mu\neq M$, we shall
follow the usual practice and set $\mu=M$.} $\mu$. To the NLO we
have
$$
C_{2+1,i}(Q,M,x,y_c,\lme)=
a(\mu/\lme)\left[c^{(1)}_{2+1,i}(Q^2,x,y_c)+
a(\mu/\lme)c^{(2)}_{2+1,i}(Q,M,x,y_c)\right],
$$
\begin{equation}
C_{1+1,i}(Q,M,x,y_c,\lme)=
c^{(0)}_{1+1,i}(Q,x)+
 a(\mu/\lme)c^{(1)}_{1+1,i}(Q,M,x,y_c).\;\;\;\;\;\;\;\;\;\;\;\;\;\;\;
\label{C1+1}
\end{equation}
In standard global analyses of hard scattering processes
\cite{MRS,CTEQ,GRV},
\lm is fitted together with a set of parameters
$a_i^{(j)}$, describing distribution functions $p^{(j)}(x,M)$
of parton species $j$ at some initial scale $M_0$, usually in the
form
\be
p^{(j)}(x,M_0)=a_0^{(j)}
x^{a_1^{(j)}}(1-x)^{a_2^{(j)}}
\left(1+a_3^{(j)}\sqrt{x}+a_4^{(j)}x+a_5^{(j)}x^{3/2}\right).
\label{initial}
\ee
Writing the derivative $\mr{d}\sigma_{2+1}/\mr{d}\ln \lme$ as
\begin{eqnarray}
&&\frac{\mr{d}\sigma_{2+1}(Q^2,\lme)}{\mr{d}\ln\Lambda}=
 \nonumber \\
&&\sum_{i}\int^1_0 \mr{d}x\left[
\frac{\mr{d}f_{i}(x,M,\lme)}{\mr{d}\ln\Lambda}
a(M)c_{2+1,i}^{(1)}(x)+f_i(x,M,\lme)c_{2+1,i}^{(1)}(x)
\frac{\mr{d}a(M)}{\mr{d}\ln\lme}\right]=  \label{l1} \nonumber \\
&&a(M)\sum_{ij}\int^1_0 \mr{d}x\left[-
\int^1_0\frac{\mr{d}y}{y}f_j(y,M,\lme)P^{(0)}_{ij}(z)
a(M)c_{2+1,i}^{(1)}(x)+ \right.\label{pu} \nonumber \\
&&\hspace*{3cm}\left.\!bf_i(x,M,\lme)
a(M)c_{2+1,i}^{(1)}(x)\right],\label{l3}
\end{eqnarray}
where $z\equiv x/y$ and $P^{(0)}_{ij}(z)$ are the LO branching
functions
\footnote{In \reff{l3} we have for brevity suppressed
the dependence of $c^{(1)}_{2+1,i}$ on $y_c$ and written $a(M/\lme)$
simply as $a(M)$.}, we see that the leading order term of
$\mr{d}\sigma_{2+1}/\mr{d}\ln \lme$ gets contributions from
the variation of \lm in both the PDF $f_i$ and
hard scattering cross--sections $C_{2+1,i}$.
The two terms in the brackets of \reff{l3} are of the
same order and their relative importance thus is basically a LO effect.

For the quantity $R_{2+1}(Q^2)$ the situation is less obvious. PDF
appear in both the numerator and denominator of \reff{R2+1} and so some
cancellations might occur, while the hard scattering cross--sections
start as $O(a)$ for $C_{2+1,i}$ and as $O(1)$ for $C_{1+1,i}$.
Nevertheless a detailed analysis \cite{preprint} shows that also for the
ratio
$R_{2+1}$ varying \lm in PDF generates terms which are of the same order
as those resulting from the variation of \lm in the hard scattering
cross-sections $C_{k+1,i}$.

\section{Parton distribution functions for arbitrary \lm}
To assess the potential of \reff{R2+1} for
a precise determination of $\alpha_s$ we wish to separate the question
of its sensitivity to $\alpha_s$ from the sensitivity to
parameters describing the initial condition on the PDF. Moreover,
in order to investigate the numerical importance of varying \lm
in PDF we have to know what to do with the parameters
$M_0,a^{(j)}_i$ in \reff{initial} when \lm is varied.
We cannot keep them fixed as this would contradict
the idea of factorization.
To see why, let us first consider the couplant as well as the
evolution equation for the nonsinglet quark distribution function
$q_{\mr{NS}}(x,M)$, at the LO. In terms of conventional moments
we have
\be
\frac{\mr{d}q_{\mr{NS}}(n,M,\lme)}{\mr{d}\ln M}=a(M/\lme)
q_{\mr{NS}}(n,M)P^{(0)}_{\mr{NS}}(n),\;\;
d_n\equiv -P^{(0)}_{\mr{NS}}(n)/b,
\label{LLmom}
\ee
whence
\be
q_{\mr{NS}}(n,M,\lme)=
A_{n}\left[a(M/\lme)\right]^{d_n},
\;\;\;a(M/\lme)=\frac{1}{b\ln (M/\lme)}.
\label{LLsolution}
\ee
In the above expressions
$P^{(0)}_{\mr{NS}}(n,M)$ are moments of LO nonsinglet branching
function and $A_n$ are \underline{unique} dimensionless constants,
describing the nonperturbative properties of the nucleon, introduced
in \cite{Politzer}.
Via \reff{LLsolution} they determine the asymptotic
behavior of $q_{\mr{NS}}(n,M,\lme)$ for $M\rightarrow \infty$
 and thus provide alternative way of specifying the
initial conditions on the solutions of \reff{LLmom}.
The separation in \reff{LLsolution} of $q_{\mr{NS}}(n,M,\lme)$ into two
parts -- one calculable in perturbation theory and the other
incorporating all the nonperturbative effects -- is the very
essence of the factorization idea. Note that in
\reff{LLsolution} the dependence on \lm is simply a reflection of its
dependence on $M$. Knowing the latter, we know the former.
Eq. \reff{LLsolution} implies that we can write down the
equations for the $\Lambda$--dependence of PDF
which are very similar to standard evolution
equations, describing their dependence on the factorization scale
$M$. The situation is basically the same as for the running couplant
$a(M/\Lambda)$ for which $\mr{d}a(M/\Lambda)/\mr{d}\ln \Lambda=
-\mr{d}a(M/\Lambda)/\mr{d}\ln M$. In the case of PDF the only
difference stems from the necessity to properly specify the initial
conditions on the solution of \reff{LLmom}.
For finite initial $M_0$ \reff{LLsolution} implies
\be
q_{\mr{NS}}(n,M,\lme)=q_{\mr{NS}}(n,M_0,\lme)
\left[\frac{a(M/\lme)}{a(M_0/\lme)}\right]^{d_n}=
q_{\mr{NS}}(n,M_0,\lme)\exp(-d_n s),
\label{MM0}
\ee
where
\be
s\equiv\ln\left(\frac{\ln(M/\lme)}{\ln(M_0/\lme)}\right)
\doteq
\ln\left(\frac{a(M_0/\lme)}{a(M/\lme)}\right)
\label{s}
\ee
is the so called ``evolution distance'' \cite{GRV}. The second equality
in \reff{s} holds exactly at the leading order and approximately at
higher ones. Rewriting \reff{MM0} again in the form \reff{LLsolution}
\be
q_{\mr{NS}}(n,M,\lme)=\left[\frac{q_{\mr{NS}}(n,M_0,\lme)}
{\left(a(M_0/\lme)\right)^{d_n}}\right]
\left(a(M/\lme)\right)^{d_n}
\label{ratio}
\ee
we see that the ratio of $q_{\mr{NS}}(n,M_0,\lme)$ and
$(a(M_0/\lme))^{d_n}$ equals $A_n$ and must therefore
be both $M_0$ and \lm
independent. This, however, is possible only if the initial moments
$q_{\mr{NS}}(n,M_0,\lme)$ do, as indicated, depend beside $M_0$ on \lm
as well! Note that considering $q_{\mr{NS}}(n,M,\lme)$ as a function of
$s$, formula \reff{ratio} implies that entire dependence of
$q_{\mr{NS}}(n,M,\lme)$ on \lm comes solely from the term
$\ln(M/\lme)=1/a(M/\lme)$ in \reff{s}! The same happens in the realistic
case of coupled quark and gluon evolution equations \cite{Buras}.
If we wish to investigate the dependence of PDF solely on \lm, keeping
fixed all other parameters, specifying the initial conditions on PDF,
we cannot fix initial conditions at some finite $M_0$, (i.e. the
parameters $p^{(j)}(x,M_0)$ in \reff{initial}). It is the
constants $A_n$ that must be kept fixed instead! On the other hand, in
global analyses, like \cite{MRS,CTEQ,GRV}, all parameters,
including \lm, are varied simultaneously and fitted to data. In these
circumstances it is then not straightforward
to determine the sensitivity of a
given physical quantity to $\alpha_s$ itself.

The preceding paragraph also suggests a simple procedure,
which makes use of some of the available parameterizations and
generalizes them to arbitrary \lm, keeping the
constants $A_n$ fixed. From all the available parameterizations of PDF
those given by analytic expressions of the coefficients $a^{(j)}_i$ on a
general factorization scale $M$ via the variable $s$
are particularly suitable for this purpose. In our studies we
took several of the GRV parameterizations, both LO and NLO ones, defined
in \cite{GRV2}. At the LO the recipe for  construction of PDF for
arbitrary \lm is simple:
\bit
\item Use any of these parameterizations with fitted value of
$\Lambda_{fit}$.
\item Keep $\lme$ in $\ln(M_0/\lme)$ fixed at the value $\Lambda_{fit}$.
\item Vary $\lme$ in $\ln(M/\lme)$.
\item Use the original parameterization with the redefined $s=s(\lme)$.
\eit
This construction is exact at the LO. At the NLO the couplant
$a(M/\lme)$ is no longer given simply as $1/\ln(M/\lme)$ and
consequently  the second equality in \reff{s} holds only approximately.
This, together with the additional term appearing in the NLO expression
for $q_{\mr{NS}}(n,M,\lme)$
\be
q_{\mr{NS}}(n,M,\lme,A_n)=
A_{n}\left[\frac{a(M/\lme)}
{1+ca(M/\lme)}\right]^{d_n}
\left(1+ca(M/\lme)\right)^{P^{(1)}_{\mr{NS}}(n)/bc},
\label{NLLsolution}
\ee
destroy simple dependence of $q_{\mr{NS}}$ on $M,M_0$ and $\lme$ via the
variable $s$. Using the above recipe for the NLO GRV parameterizations
provides therefore merely an approximate solution of our task, but one
that is sufficient for the purposes of determining the relative
importance of varying \lm in PDF and hard scattering cross--sections
as this relative importance is basically a LO effect.
\begin{figure}
\epsfig{file=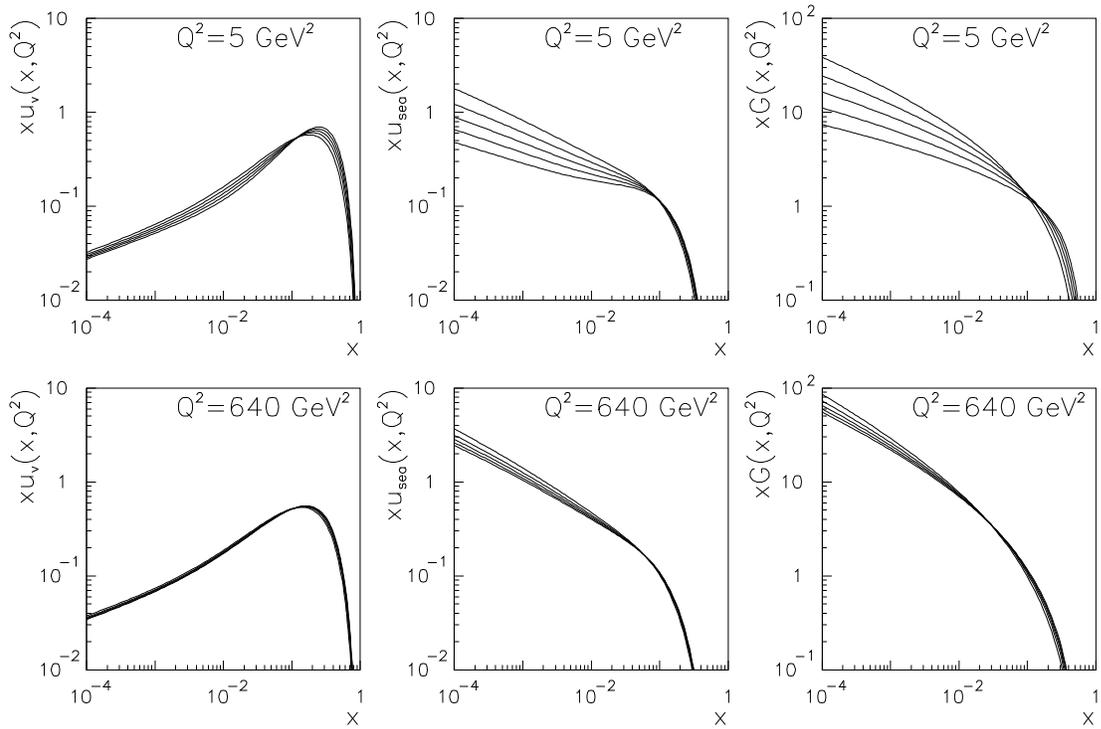,width=15cm}
\caption{The dependence of $xu_v(x,Q^2), xu_{\mr{sea}}(x,Q^2)$ and
$xG(x,Q^2)$ of the value $\lmepdf$ according to the recipe described
in the text and for two values of $Q^2$. At low $x$, the curves
correspond from above to the values of $\lmepdf$, subsequently,
$0.1,0.2,0.3,0.4$ and $0.5$.}
\end{figure}

As an illustration of our procedure we plot in Fig. 1, for $x\ge
10^{-4}$ and two values of $Q^2$, the dependence of valence and sea
$u$--quark and gluon distribution functions on $\lme$. The curves
correspond to the LO GRV parameterization of \cite{GRV2}.
We see that this
dependence decreases with increasing $Q^2$, is most pronounced for the
gluon distribution function and almost irrelevant for the valence quark
one. These features are qualitatively the same as those obtained in
refs. \cite{Andreas,Alan}, which contain results of global fits
performed for several fixed values of $\lme$. Quantitatively, however,
the sensitivity to the variation of \lm, observed in these papers, is
markedly weaker than that displayed in Fig. 1. This is due to the fact
that our $\Lambda$--dependent PDF \underline{are not} constructed to
describe the hard scattering data for all values of \lm
(they do so by definition only for $\Lambda=\Lambda_{fit}$)
but in order to facilitate studies of the dependence of physical
quantities solely
on \lm, with the appropriate parameters specifying the initial
conditions kept fixed. On the other hand, in
global analyses of \cite{Andreas,Alan} fitting the data for different
values of $\lme$ leads
to different values of boundary condition
parameters $a^{(j)}_i$ at $M_0$,
which partially compensates the variation of $\lme$.

\section{Numerical results}
All the results reported below were obtained using the PROJET 4.1
generator \cite{PROJET}. It is true that the JADE jet algorithm used
in this program has certain theoretical shortcomings, but as the H1
Collaboration used it, we did the same.
Although  MC generator MEPJET \cite{MEPJET} is to be
preferred on theoretical grounds, the main features of our results are
unlikely to change because the $\Lambda$-dependence of $R_{2+1}$ is
basically a LO effect.  To quantify the dependence of $R_{2+1}$ on \lm
separately in hard scattering cross--sections and PDF we consider it
as a function of $Q^2, y_c$ and two independent $\Lambda$--parameters,
denoted as $\lme^{\mr{hsc}}$ and $\lme^{\mr{pdf}}$ respectively. The
results are then studied as a function of $\lmehsc,
\lmepdf, Q^2$ and $y_c$ for
\bit
\item 12 values of $Q^2$ (equidistant in $\ln Q^2$) between 5 and
$10^4$ GeV$^2$,
\item 5 values of $\lmepdf$ or $\lmehsc$, equal to $0.1,0.2,0.3,0.4$
and $0.5$ GeV,
\item 3 values of $y_c=0.01,0.02,0.04$.
\eit

\subsection{No cuts on jets}
First we analyse the case with no cuts on the final state jets and
then discuss the changes caused by the imposition of cuts as specified
in \cite{H1}.
\begin{figure}
\begin{center}
\epsfig{file=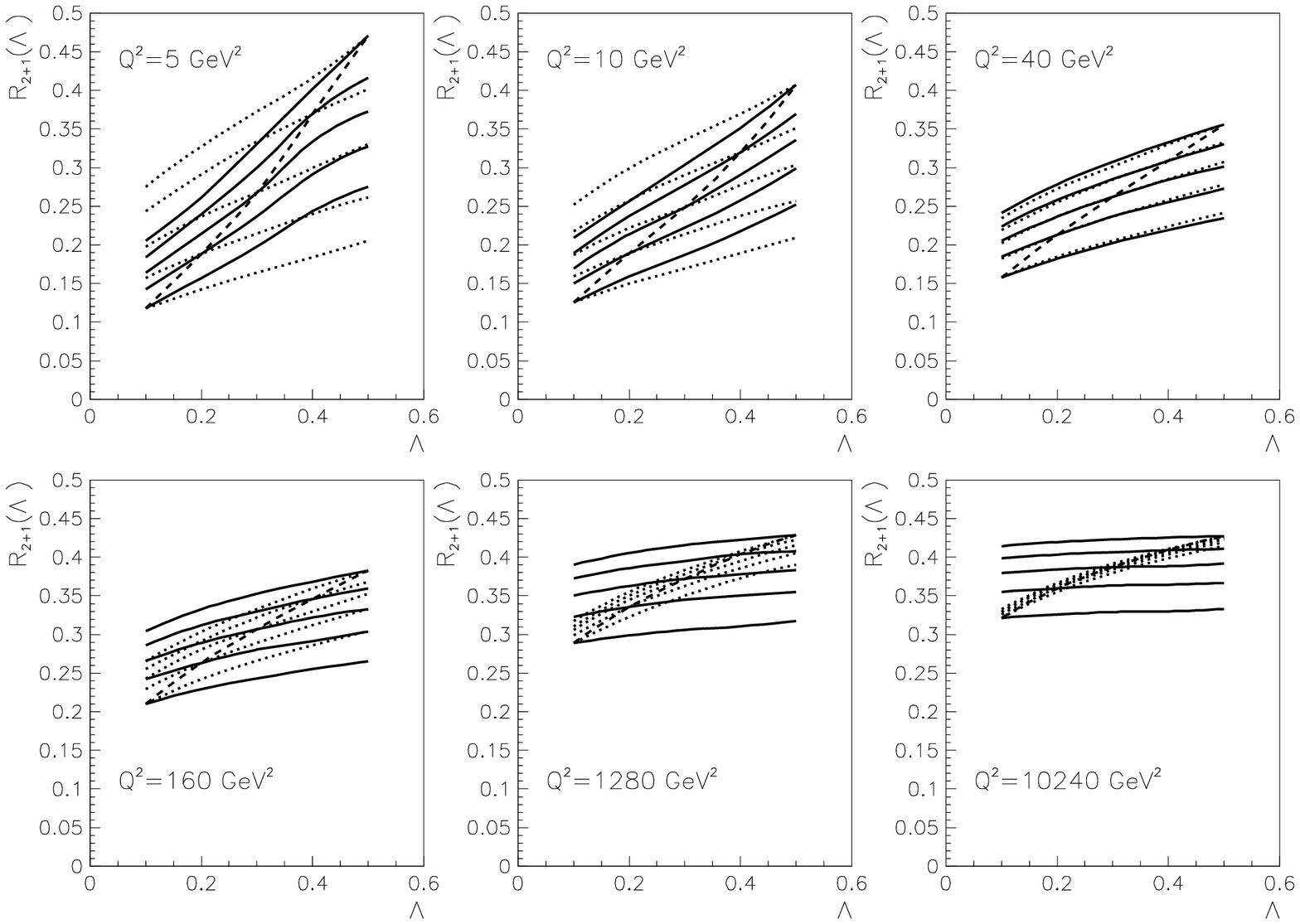,width=15cm}
\end{center}
\caption{The ratio $R_{2+1}(Q,y_c,\lmepdf,\lmehsc)$ as a function of
$\lmepdf$ for fixed $\lmehsc$ and a series of $Q^2$ values (solid
lines), or as a function of $\lmehsc$ for fixed $\lmepdf$ and the same
set of $Q^2$ values (dotted lines). The dashed lines correspond to
simultaneous variation of $\lmepdf=\lmehsc$. Solid curves
correspond, from below, to fixed $\lmehsc=0.1,0.2,0.3,0.4,0.5$ and the
dotted to the same fixed values of $\lmepdf$. In all plots $y_c$
equals $0.01$.}
\end{figure}
Fig. 2 shows a series of plots corresponding to $y_c=0.01$ and
displaying the dependence of $R_{2+1}^{\mr{NLO}}$ on $\lmehsc$ for
fixed $\lmepdf$
(dotted curves) or on $\lmepdf$ for fixed $\lmehsc$ (solid curves).
Each plot corresponds to one of the 6 selected values of $Q^2$.
We see that for $Q^2$ below about $40$ GeV$^2$ the solid curves are
steeper that the dotted ones, while above $40$ GeV$^2$ the situation is
reversed. This means that below $40$ GeV$^2$ varying \lm in PDF is more
important than varying it in the hard scattering cross--sections for
$Q^2$, while the opposite holds above $40$ GeV$^2$. The same
information as in Fig. 2 is displayed in a different manner
in Fig. 3, where the $Q^2$ dependence of $R_{2+1}$ is plotted for
various combinations of $\lmepdf$ and $\lmehsc$. The shape of the curves
in Fig. 3 is a result of two opposite effects: as $Q^2$ increases the
phase space available for produced jets increases as well, but at the
same time the value of the running $\alpha_s(Q/\lme)$ decreases. At low
$Q^2$, on the other hand, phase space shrinks, but $\alpha_s(Q/\Lambda)$
grows.
\begin{figure}
\begin{center}
\epsfig{file=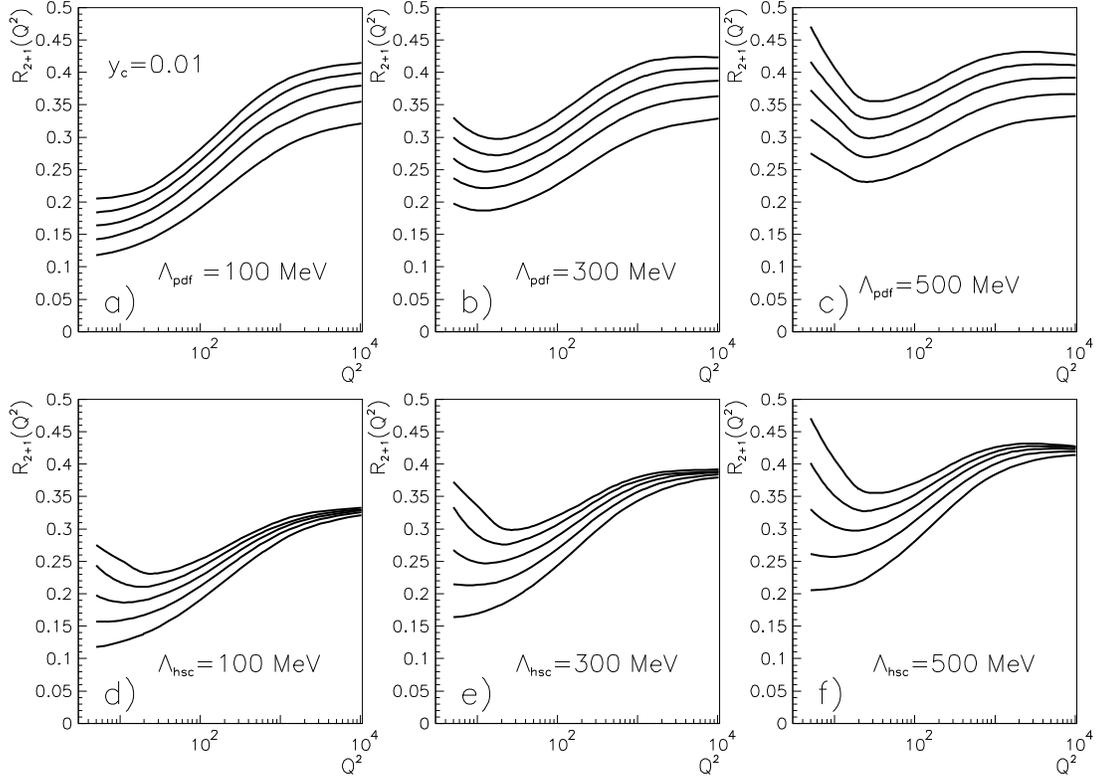,width=15cm}
\end{center}
\caption{(a--c) $Q^2$--dependence of the ratio $R_{2+1}(Q,y_c,\lmepdf,\lmehsc)$
for fixed $\lmepdf=0.1,0.3,0.5$ GeV and five values of
$\lmehsc=0.1,0.2,0.3,0.4,0,5$ GeV. (d--f) The role of
$\lmepdf$ and $\lmehsc$ is reversed. No cuts were applied in
evaluating $R_{2+1}$ and $y_c=0.01$. In all plots the curves
are ordered from below according to increasing $\lmehsc$ (in a--c) or
$\lmepdf$ (in d--f).} \end{figure}

In Fig. 3 the relative importance of varying $\lmepdf$ and $\lmehsc$
is reflected in a bigger spread of the five curves in Figs. 3d)-f)
compared to those in Fig. 3a)-c) for $Q^2\le 40$ GeV$^2$ and smaller
spread above $40$ GeV$^2$. Note also that for fixed $\lmepdf$ the
sensitivity to the variation of $\lmehsc$ is about the same at all
$Q^2$. The shape of curves in Fig. 2 suggests that the ratio
\be
V(Q^2,y_c,r,s)\equiv
     \frac{R_{2+1}(Q^2,y_c,\lmepdf=s,\lmehsc=r)}
{R_{2+1}(Q^2,y_c,\lmepdf=r,\lmehsc=s)},
\label{V}
\ee
is approximately a linear function of $s$ for any fixed
$Q^2,y_c,r$. Considered as a function of $s$ for fixed $r$,
\reff{V} quantifies the relative importance of varying \lm in
PDF and hard scattering cross--sections.  To
summarize the results of Fig. 2 we fitted $V(Q^2,y_c,r,s)$
by a linear function of $s$ and defined
\be
W(Q^2,y_c,r)\equiv \frac{\mr{d}V^{fit}(Q^2,y_c,r,s)}{\mr{d}s},
\label{W} \ee
which, by construction, is $s$--independent function of $Q^2,y_c$ and
$r$. Positive $W$ means that the variation of \lm in the PDF is more
important than that in the hard scattering cross--sections, while
for negative $W$ the situation is opposite.
\begin{figure}
\begin{center}
\epsfig{file=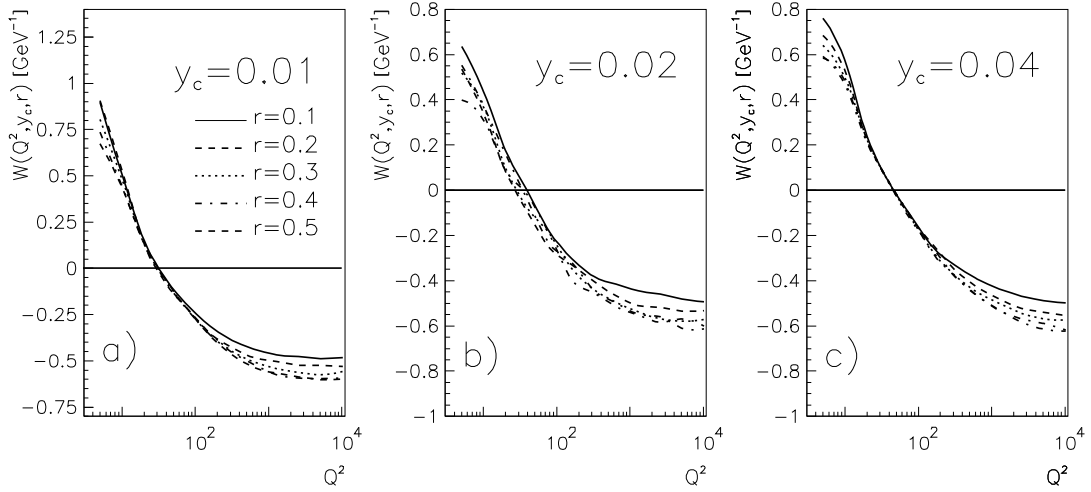,width=15cm}
\end{center}
\caption{(a--c) The quantity $W(Q^2,y_c,r)$ as a function of $Q^2$ for five
values of $r$ (given in GeV) and $y_c=0.01,0.02,0.04$. All curves
correspond to the case of no cuts.}
 \end{figure}
The $Q^2$--dependence of $W(Q^2,y_c,r)$ is plotted in
Fig. 4a-c for three values of $y_c=0.01,0.02,0.04$ and five values of
$r=0.1,0.2,0.3,0.4,0.5$ GeV. In Fig. 5a we compare the $Q^2$ dependence
of $W(Q^2,y_c,r=0.2\;\mr{GeV})$ for different values of
$y_c=0.01,0.02,0.04$. Similar plots can be drawn for other values of $r$
as well. From Figs. 2--5a we conclude that at the NLO:
\bit
\item For $Q^2$  below $40$ GeV$^2$ the sensitivity of \reff{R2+1} to
$\lme$ comes dominantly from PDF, while above $40$ GeV$^2$ the situation
rapidly changes and most of this sensitivity comes from hard scattering
cross--section.
\item
Above $Q^2\approx 10^3$ GeV$^2$ the sensitivity to \lm in PDF becomes
negligible.
\item The preceding conclusions depend only weakly on $y_c$.
\eit
\begin{figure}
\begin{center}
\epsfig{file=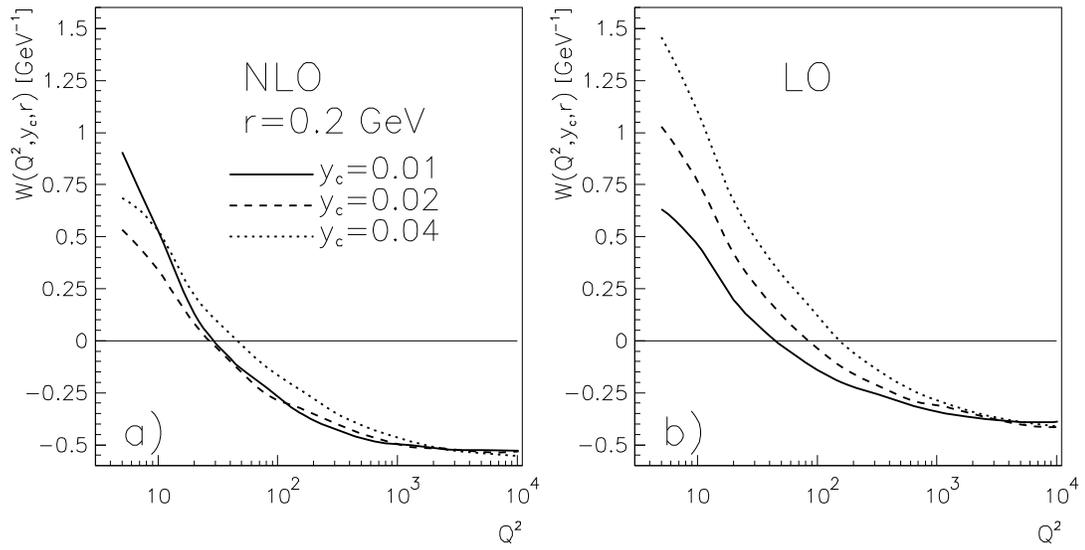,width=15cm}
\end{center}
\caption{The $Q^2$--dependence of $W(Q^2,y_c,r=0.2)$ for three
different values of $y_c$ at the NLO (a) and LO (b). All curves
correspond to the case of no cuts.}
\end{figure}
At the LO the general features are the same as those displayed in Figs.
2--5a and we therefore merely summarize them in Fig. 5b. Comparing
Figs. 5a and 5b we conclude that at the LO the relative importance of
varying \lmpdf with respect to \lmhsc is bigger than at the NLO.

\subsection{H1 cuts}
In \cite{H1} only events satisfying certain acceptance cuts, most
notably the cut on polar angle of outgoing jets in HERA
laboratory frame,
$10^{\circ}\le\theta^{\mr{jet}}_{\mr{lab}} \le 145^{\circ}$, were
selected for the analysis.
\begin{figure}
\begin{center}
\epsfig{file=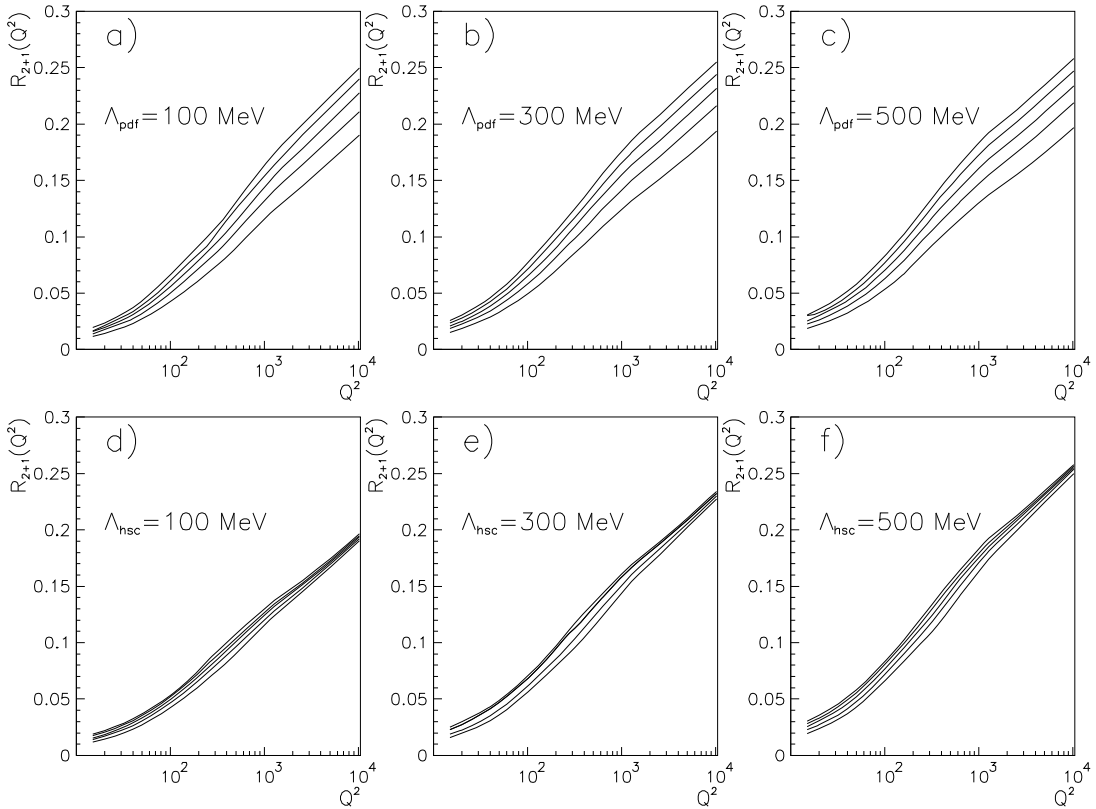,width=15cm}
\end{center}
\caption{The same as in Fig. 3 but for the H1 acceptance cuts.}
\end{figure}
In Fig. 6 plots analogous to those of Fig. 3, but incorporating these
cuts, are displayed. We see that the experimental cuts affect
mostly the low $Q^2$ region, where
beside lowering the values of $R_{2+1}(Q^2)$ they also
strongly suppress the
sensitivity of $R_{2+1}(Q^2)$ to the variation of \lm in the PDF.
This effect can be understood as follows. By imposing
the mentioned angular cuts, one removes events with
jets that fly into a large part
of the backward (with respect to the proton direction)
lab frame hemisphere.
We recall that
in direct photon interactions (in contrast to the resolved ones), i.e.
also DIS, a large fraction of jets populates the backward hemisphere
in the $\gamma$p CMS.  Because the transverse boost
between the $\gamma$p CMS and HERA lab frame is proportional to
$Q^2$, the cut on the jet polar angle in the lab system is more
effective for small $Q^2$ than for high $Q^2$.
Hence the drastic suppression of $R_{2+1}(Q^2)$ at low and
only moderate at high $Q^2$.
Furthermore, as low $Q^2$ means in average also small $x$, and
most of the sensitivity of $R_{2+1}(Q^2)$ to $\Lambda$ in PDF comes
from the small $x$ region, the H1 cuts
will significantly reduce it mainly for low $Q^2$.

On the basis of Fig. 6 we conclude that
in the region $Q^2\ge 100$ GeV$^2$, used in
\cite{H1} for the extraction of \lm from the measured $R_{2+1}(Q^2)$,
the variation of \lm in the PDF could be neglected with respect to
the variation of $\Lambda$ in the hard scattering cross--sections.

\section{Conclusions}
In this paper we have constructed parameterizations of PDF corresponding
to solutions of the evolution equations with fixed suitably defined
boundary conditions and variable \lm. These
parameterization were then used to investigate, for the quantity
$R_{2+1}$, the  numerical importance of varying \lm in PDF. We
concluded that:
\bit
\item
If no cuts are applied on produced jets the sensitivity of \reff{R2+1}
to variation of \lm in PDF
is bigger at the LO than at the NLO.
\item At moderate $Q^2$, roughly below $40$ GeV$^2$, the sensitivity
of \reff{R2+1} to $\alpha_s$ comes dominantly from PDF, while at higher
$Q^2$ the hard scattering cross--sections take rapidly over as the
main source of the $\Lambda$--dependence of \reff{R2+1}.
\item The preceding conclusions are only weakly dependent on $y_c$.
\item Including the cuts applied by the H1 Collaboration leads to
strong suppression of the sensitivity to $\alpha_s$ in the moderate $Q^2$
region.
\eit
It is clear that a less restrictive treatment of the jets in the
$Q^2$ region between, say, 10 and 100 GeV$^2$ could
provide much better possibility for a precise determination of
$\alpha_s$. This region is also very interesting from another point
of view, and namely the question of the transition between the
dynamics of deep inelastic scattering and photoproduction. The data
in this region could bring new information on the
structure of the virtual photon \cite{jacvach} and/or lead to
new phenomena.

\vspace*{0.3cm}
\noindent
{\large \bf Acknowledgments}

\vspace*{0.15cm}
\noindent
This work had been supported in part by the grant No. 201/96/1616 of the
Grant Agency of the Czech Republic.

\end{document}